\documentclass[preprint]{aastex}

\shorttitle{MeV Spectrum of Cygnus X-1}
\shortauthors{McConnell et al.}

\begin{document}

\title{A High Sensitivity Measurement of the MeV $\gamma$-Ray Spectrum
       of Cygnus X-1}

\author{M. L. McConnell, J. M. Ryan}
\affil{ Space Science Center,University of New Hampshire, Durham, NH 03824}
\email{Mark.McConnell@unh.edu,James.Ryan@unh.edu}

\author{W.  Collmar, V. Sch\"onfelder, H.  Steinle, A. W. Strong}
\affil{ Max Planck Institute for Extraterrestrial Physics, Garching, Germany}
\email{wec@mpe.mpg.de,vos@mpe.mpg.de,hcs@mpe.mpg.de,aws@mpe.mpg.de}

\author{H. Bloemen, W.  Hermsen, L. Kuiper}
\affil{ Space Research Organization of the Netherlands (SRON), Utrecht,The 
Netherlands}
\email{h.bloemen@sron.nl,w.hermsen@sron.nl,l.m.kuiper@sron.nl}

\author{K. Bennett}
\affil{ Astrophysics Division, ESTEC, Noordwijk, The Netherlands}
\email{kbennett@astro.estec.esa.nl}

\author{B. F. Phlips}
\affil{George Mason University, Fairfax, VA 22030}
\email{phlips@gamma.nrl.navy.mil}

\and

\author{J. C. Ling}
\affil{Jet Propulsion Laboratory, California Institute of Technology, Pasadena,
CA 91109}
\email{jling@jplsp.jpl.nasa.gov}

\begin{abstract}
The Compton Gamma-Ray Observatory
(CGRO) has observed the Cygnus region on several occasions since 
its launch in 1991.  The data collected by the COMPTEL experiment
on CGRO represent the
most sensitive observations to date of Cygnus X-1 in the 0.75-30 MeV
range. A spectrum accumulated by COMPTEL over 10 weeks of observation time
shows significant evidence for emission extending out to several MeV.
We have combined these data with contemporaneous data from both BATSE and OSSE 
to produce a broad-band $\gamma$-ray spectrum, corresponding to the low
X-ray state of Cygnus X-1, extending from 50 keV up to 
$\sim$5 MeV.  
Although there is no evidence for any broad line-like emissions 
in the MeV region, these data further confirm the presence of a hard tail
at energies above several hundred keV.
In particular, the spectrum at MeV energies can be described as a power-law 
with a photon spectral index of $\alpha$ = -3.2, with no evidence 
for a cutoff at high energies.   
For the 200 keV to 5 MeV spectrum, 
we provide a quantitative description of the underlying electron spectrum, 
in the context of a hybrid thermal/non-thermal model for the emission.
The electron spectrum can be described by a thermal Maxwellian with a 
temperature of $kT_e$ = 86 keV and a non-thermal power-law component with a 
spectral index of $p_e$ = 4.5.  The spectral data presented here should 
provide a useful basis for further theoretical modeling.

\end{abstract}


\clearpage

\section{Introduction}

Cygnus X-1 is generally considered to be one of the most well-established
candidates for a stellar-mass black hole.  Having been discovered as an
X-ray source more than 30 years ago, it has been studied extensively at
X-ray and $\gamma$-ray energies.  At energies approaching 1 MeV, it is
one of the brightest sources in the sky.  The spectrum at MeV energies
is so steep, however, that it has been poorly measured at energies near 
1 MeV and above.  An accurate characterization of the spectrum at these 
high energies will facilitate a more complete understanding of the 
underlying physics of this source. This, in turn, 
will surely have an impact on our understanding of the
$\gamma$-ray emission from all black hole sources, including the
(stellar mass) soft X-ray transients and the (supermassive) Active
Galactic Nuclei (AGN).

It has long been recognized that the soft X-ray emission ($\sim 10$ keV)
generally varies between two discrete levels \citep[e.g.,][]{priedhorsky83,ling83,liang83}.  
Cygnus X-1 seems to spend most ($\sim 90\%$) of 
its time in the so-called {\em low X-ray state}, characterized by a relatively low flux of 
soft X-rays and a relatively high flux of hard X-rays ($\sim100$ keV).  This state 
is sometimes referred to as the {\em hard state}, based on the nature of its soft X-ray
spectrum.  On occasion, it 
moves into the so-called {\em high X-ray state}, characterized by a relatively high soft 
X-ray flux and a relatively low hard X-ray flux.   
This state 
is sometimes referred to as the {\em soft state}, based on the nature of its soft X-ray
spectrum.  There are, however, some exceptions to this general behavior.  For example,  
HEAO-3 observed, in 1979,  a relatively low hard X-ray flux coexisting with a low 
level of soft X-ray flux \citep{ling83,ling87}.  \citet{ubertini91a} observed a similar 
behavior in 1987.

Early X-ray spectra of Cygnus X-1 (in its low X-ray state) 
supported the notion that the high
energy emission resulted from the accretion of matter onto a
stellar-mass black hole.  The emission was generally interpreted to be
the result of the Comptonization of a soft thermal photon flux by an
energetic electron population.  The Comptonization model of \citet[hereafter ST80]{sunyaev80}
became the standard model for the
interpretation of high energy spectra from Cygnus X-1 and other similar
sources.  This model assumed some (unspecified) source of soft photons 
interacting in an optically thick 
Comptonization region ($\tau \gg 1$) with nonrelativistic electrons 
($kT_e \ll m_{e}c^2$).  It was successfully used to interpret 
many of the early hard X-ray observations at 
energies below $\sim200$ keV. 
The data for Cygnus X-1 indicated 
that the Comptonization was taking place in a region with an electron 
temperature ($kT_e$) in the 30--60 keV range and an optical depth 
($\tau$) of 1--5 \citep[e.g.,][]{sunyaev79,steinle82,ubertini91b}.

As the observations improved, especially at energies 
extending beyond 200 keV, it became increasingly difficult to model the 
broad-band continuum spectrum with a single-temperature (or single-component)
inverse Compton model \citep[e.g.,][]{nolan81,nolan83,grebenev93}.
Additional spectral components, such as a second Comptonization region,
were invoked to improve the spectral fits.
\citet{grebenev93}
suggested that the discrepancy at higher energies could also be explained as
a result of the limitations of the ST80 model.  In particular, they
showed that Monte Carlo simulations of the Comptonization process
(which were not restricted by the assumptions of the analytical model)
could provide accurate fits to the broad-band GRANAT data extending up to 
several hundred keV.

Meanwhile, there were several efforts designed to expand upon the 
fundamentals provided by the ST80 model.  \citet{zdziarski84,zdziarski85,zdziarski86} 
considered both bremsstrahlung and synchrotron radiation as soft photon 
sources.   Further analytical improvements to the ST80 Comptonization model were 
developed by
\citet[][hereafter, T94]{titarchuk94}.  (See also \citet{hua95} and
 \citet{titarchuk95a}.)
Incorporating various relativistic corrections, 
this  so-called generalized Comptonization model
was applicable over a much wider range of parameter space.  The
ability of the T94 model to accurately predict the spectrum over a broader
range of spectral parameters has been subsequently verified via Monte
Carlo simulations by both \citet{titarchuk95b} and by \citet{skibo95a}.  
The simulations of \citet{skibo95a} included both bremsstrahlung and annihilation  
radiation as soft photon sources, in addition to some (unspecified) external 
soft photon source.  They concluded that the ST80 model agreed 
well with Monte Carlo simulations
for $kT_e \lesssim 200$ keV and $\tau \gtrsim 2$, whereas the T94 model agreed 
well with Monte Carlo simulations for $kT_e \lesssim 300$ keV and $\tau \gtrsim 0.2$.
These same simulations also demonstrated that, under certain conditions, the 
analytical model of \citet{zdziarski85} could also be used.

Source geometry is also an important factor that must be considered in the 
interpretation of broad-band spectra.  For example, \citet{haardt93} argued, 
based on Monte Carlo studies, that improved fits
to broad-band spectra could be achieved by incorporating a 
reflection component along with the inverse 
Compton component. The reflection component would result from the
Compton scattering of hard X-rays from an accretion disk corona off a cool
optically-thick accretion disk.  The reprocessing leads to an enhancement of  
emission in the 10--100 keV energy range.  The requirement for such a 
reprocessed component is also supported by the observation of a fluorescence line from 
neutral iron \citep[e.g.,][]{ebisawa96}.  
Assuming a geometry consisting of an optically thin corona above an optically-thick 
accretion disk, \citet{haardt93} showed that the model was consistent 
with data from EXOSAT, SIGMA \citep{salotti92}
and OSSE \citep{grabelsky93}, suggesting a much higher coronal electron
temperature ($kT_e \sim 150$ keV) and an even smaller optical depth
($\tau \sim 0.3$) than suggested by Comptonization models alone.  
The reflection component was also incorporated into an accretion disk corona (ADC) model 
by \citet{dove97b}.  In this case, the geometry consisted of a hot inner spherical corona
with an exterior accretion disk. 
Reasonable comparisons with the broad-band spectrum from 1 keV up to several hundred keV
were demonstrated with an electron temperature of $kT_e \sim 90$ keV and an optical depth of $\tau \sim 1.5$. 
A similar model was used by \citet{gierlinski97} to fit a combined Ginga-OSSE spectrum, but 
they found that, with fixed normalization between the Ginga and OSSE spectra,
a second Comptonization component was needed to improve the fit at the 
highest energies.  (It has been pointed out by \citet{poutanen2000}, however, 
that a single temperature Comptonization 
model fits the Ginga-OSSE data quite well if the relative normalization between the two 
spectra is left as a free parameter;
see Figure 6 of \citet{poutanen98a}.)
\citet{moskalenko98} also developed 
a multi-component model to explain the spectrum from soft X-rays into 
the $\gamma$-ray region.  In this case, a central spherical corona surrounds the 
black hole itself, outside of which is the optically-thick accretion disk, with a 
much hotter outer corona surrounding the entire system.

The use of two-component Comptonization models to improve spectral fits 
is motivated by the reasonable assumption that 
the Comptonization region will not be isothermal.  A simple two-temperature model may not be very 
physical, however, since it implicitly assumes that there are two distinct non-interacting 
regions in which Comptonization is taking place.  A more realistic approach would be 
to assume a continuum of Comptonization parameters.
\citet{skibo95b} argued that a
thermally-stratified black hole atmosphere could act to harden the
spectrum.  They simulated a model involving a high-energy spherical core 
surrounded by an optically thick accretion disk.  
If the inner core of the high-energy region
were hot enough, then a hard tail extending into the MeV region might
be produced.  A similar model was used to interpret the first combined 
spectral results from the BATSE and COMPTEL experiments on CGRO \citep{ling97} .  
Thermal stratification is also an essential concept in the transition disk 
model of \citet{misra96}.  This model involves large thermal gradients in the 
inner region of an optically thick accretion disk.  These gradients
represent a transition between the cold optically thick disk and the hot plasma 
which exists near the inner part of the disk.  This model has been used to provide 
a good fit to spectra in the 2--500 keV energy band \citep{misra97,misra98}.
\citet{dove97b} also incorporated thermal stratification into the structure of 
the inner corona of their model.

The net effect of thermal gradients is to produce a non-Maxwellian electron 
energy distribution.  A high energy tail in the electron distribution  
leads, via the the inverse Compton process, to a high energy tail in the 
photon distribution.  
Several models
have sought to explain hard-tail emissions by invoking some kind of  nonthermal process
to generate a high energy (possibly relativistic) tail to the thermal 
Maxwellian electron energy distribution.  
Many of these hybrid thermal/non-thermal models 
\citep[e.g.,][]{bednarek90,crider97,gierlinski97,poutanen98a,poutanen98b,coppi99} 
tend to be rather ad hoc, in that they assume
some accelerated (or non-Maxwellian) particle population and then proceed
to explore the subsequent consequences.  Typically, although not always, the 
nonthermal population takes the form of a power-law in the electron energy distribution.
Such a distribution is similar to that seen, for example, in solar flares \citep[e.g.,][]{coppi99}.
Both \citet{crider97} and \citet{poutanen98b} have shown that a Maxwellian plus 
power-law form for the electron energy distribution can 
be adapted to fit a composite COMPTEL-OSSE spectrum of Cygnus X-1.

Others have considered physical mechanisms
by which non-thermal electron distributions might be developed.  
For example, both stochastic particle acceleration \citep{dermer96,li96} and MHD turbulence 
\citep{li97} have been proposed as mechanisms for directly accelerating the 
electrons.  The ion popoulation might also contribute to the non-thermal 
electron distribution in the case where a two-temperature plasma develops
\citep[e.g.,][]{dahlbacka74,shapiro76,chakrabarti95}.  
In this situation, the ion population may reach a
temperature of $kT_i \sim 10^{12}$ K, resulting in $\pi^o$ production
from proton-proton interactions \citep[e.g.,][]{eilik80,eilik83}.
The $\pi^o$ component then leads, via photon-photon
interactions between the $\pi^o$-decay photons and the X-ray photons,
to production of energetic (nonthermal) $e^{+}-e^{-}$ pairs. 
\citet{jourdain94} used this concept to 
fit the hard X-ray tails of not only Cygnus X-1, but also
GRO J0422+32 and GX 339-4, as measured by both SIGMA and OSSE. 
While retaining the standard
ST80 spectrum to explain the emission at energies below
200 keV, they used  $\pi^o$ production to generate the nonthermal 
pairs needed
to fit the spectrum at energies above $\sim 200$ keV.

The history of Cygnus X-1 is also riddled with unconfirmed reports of very 
intense, very broad line emission in
the region around 1 MeV, far exceeding that which would be
expected from a simple extrapolation of the low energy continuum spectrum 
\citep[e.g.,][]{ling87,mcconnell89,owens92}.
The broad MeV feature observed by HEAO-3 \citep{ling87} occured
under an unusual source condition when both the hard and soft X-ray fluxes 
were low.  
\citet{liang88} interpreted the feature as evidence for the presence of a very hot
($kT_e \sim 400$ keV) pair-dominated cloud in the inner region of the accretion 
disk.  In this model, pairs may escape the system and produce a weak narrow
annihilation feature in the cold surrounding medium \citep{dermer89}.
Such a feature was also suggested in the HEAO-3 spectrum \citep{ling89}.
\citet{melia93} extended this 
model by considering a more realistic thermally stratified cloud.
These ``MeV bumps,'' 
have not been seen by any of the experiments on the Compton Gamma-Ray 
Observatory (CGRO)
\citep{mcconnell94,phlips96,ling97}.  
Any such emission must therefore be time-variable \citep[c.f.,][]{harris93}.  

Although there have been no recent observations of an ``MeV bump'' in the spectrum 
of Cygnus X-1, the continuum flux levels that are 
now generally observed around 1 MeV still indicate a substantial hardening 
of the high energy spectrum.  
The extent to which the spectrum hardens at energies approaching 1 MeV 
has now become an important issue for theoretical modeling of 
the spectrum.
Data from CGRO, in particular the 
COMPTEL experiment on CGRO, offer the best opportunity for more precisely
defining the highest energy parts of the spectrum.  Such 
measurements are critically important in our efforts to
determine the nature of the complete hard X-ray / $\gamma$-ray spectrum.

In this paper, we present a composite low X-ray state spectrum of Cygnus X-1 compiled from 
contemporaneous CGRO observations that spans the energy range from $\sim$30 keV 
up to $\sim 5$ MeV.  This spectrum should add new insights into the 
modeling of the broad band $\gamma$-ray spectrum of Cygnus X-1 and especially its 
emissions just above 1 MeV.  In $\S$ 2 we briefly describe the COMPTEL experiment,
the data from which have been the focus of this investigation.  The observations and 
data selection are described in $\S$ 3. In $\S$ 4 we discuss the analysis of the 
COMPTEL data.  The analysis is expanded in $\S$ 5 to include data from both OSSE 
and BATSE. A discussion of the results is then presented in $\S$ 6.

\section{The COMPTEL Experiment}

The COMPTEL experiment images 0.75--30 MeV $\gamma$-radiation within a
field-of-view of $\sim 1$ sterdian.  It consists of two independent
layers of detector modules separated by 150 cm.  The upper (D1) layer
is composed of 7 independent NE213A liquid scintillator detectors,
each 28 cm in diameter and 8.5 cm thick.  The lower (D2) layer is
composed of 14 independent NaI(Tl) detectors, each 28 cm in diameter
and 7.5 cm thick. An event is defined as a coincident interaction in a
single D1 detector and a single D2 detector.  For each event, the
total energy is estimated as the sum of the energy losses in both D1
and D2.  The interaction locations in both detectors are determined
using the relative responses of the various PMTs which are affixed to
each D1/D2 detector.  In this manner, the interaction locations can be
determined with an uncertainty of about 2.0 cm.  The measurement of
the time-of-flight (TOF) between D1 and D2 and the pulse-shape (PSD)
in D1 allow for the rejection of a large fraction of background
events, including both upward-moving and neutron-induced events.  A
more detailed description of COMPTEL can be found in 
\citet{schonfelder93}.

The basic principle of COMPTEL imaging is governed by the physics of Compton
scattering.  If the energy loss of the Compton scattered electron is 
completely contained within the D1 module and if the scattered photon is
completed absorbed by the D2 module, then we have an accurate measure of 
both the scattered
electron energy and the scattered photon energy.  Knowledge about the
interaction locations in both the upper (D1) and lower (D2) detector
layers provides information about the path of the scattered
photon.  Without more complete knowledge regarding the direction of
the scattered electron, the possible arrival direction of the incident photon
is confined to a annular region on the sky.  This {\sl event circle} is
defined to have an angular radius ($\bar{\phi}$) determined by the
Compton scattering formula:

\begin{equation}
\cos \bar{\phi} = 1 - {m_e c^2 \over E_2} + {m_e c^2 \over {E_1 + E_2}},
\end{equation}

\noindent where $m_e c^2$ is the electron rest mass energy and $E_1$
and $E_2$ are the energy deposits measured in D1 and D2, respectively.
The superposition of many event circles can lead to the
crude localization of a source by determining the direction in which
the majority of the event circles intersect (c.f., Figure 7 of Winkler
et al. 1992).  In practice, the analysis of COMPTEL data is far more
complex.  Many instrumental effects complicate this simplified
approach.  For example, incomplete energy absorption in either D1 or
D2 can render equation (1) invalid.  Effects such as these are taken
into account in the final analysis via an appropriate instrumental 
point spread function (PSF).

\section{The Observations}

Since the launch of CGRO in April of 1991, several observations of the 
Cygnus region have been carried
out with COMPTEL.  In order to assemble a broad-band picture 
of the $\gamma$-ray emission from Cygnus X-1, we set out to combine
COMPTEL data with data from the BATSE and OSSE experiments on CGRO.
The BATSE experiment is an uncollimated array of NaI scintillation 
detectors covering $4\pi$ steradian and operating in the 
$\sim$30 keV -- 1.8 MeV energy range. The spectral 
analysis of a given point source is  achieved using 
techniques that rely on the occultation of the source flux by the 
Earth.  There are currently two approaches to spectral analysis with 
BATSE.  The BATSE experiment team routinely uses a technique that 
uses data only from a limited time period about each Earth occultation
(both ingress and egress; \citet{harmon92}).
Another approach, developed by \citet{ling96}, 
makes use of a more extensive set of data in an 
effort to model the combined effect of some 65 celestial sources of 
$\gamma$-radiation and various models of the instrumental background.
This method of BATSE source analysis is embodied in 
a software system known as the Enhanced BATSE Occultation Package 
\citep[EBOP;][]{ling96, ling2000}.
The OSSE experiment is a collimated array of NaI detectors that are 
used for both on-source and off-source measurements.  OSSE operates 
in the 50 keV -- 10 MeV energy range.  

In selecting those CGRO observations to use in our broad-band analysis, we
required contemporaneous observations by all three instruments (COMPTEL, 
BATSE and OSSE).  The COMPTEL FoV is centered on the CGRO pointing 
direction (the z-axis) and can generally study sources that are within 
$40\arcdeg$ of the pointing direction.  OSSE has a more
limited FoV of $11\arcdeg$ by $4\arcdeg$ and is restricted to pointing directions 
along the spacecraft X-Z plane. BATSE, on the other hand, can observe 
Cygnus X-1 continually by means of Earth occultation techniques.      
The data selection was therefore  driven by the availability of
contemporaneous COMPTEL and OSSE data.  The selection was also confined to
the first three cycles of CGRO observations (1991 -- 1994), based on 
the quality of the COMPTEL data that was available at the time this study
was undertaken.
The initial selection of CGRO observation periods, based on these 
criteria, is
listed in Table 1.  In addition to the dates of each observation, the
table also gives the COMPTEL viewing angle (the offset angle
from the COMPTEL pointing direction) and a crude estimate of COMPTEL's total
effective on-axis exposure to Cyg X-1 (measured in days).  The exposure estimate is based
on an approximated angular response for COMPTEL and includes estimates
for the effects of Earth occultation times and telemetry contact times.

The initial analaysis of the broad-band spectrum produced from the full 
set of observations in Table 1 led to a significant discrepancy 
between the BATSE and OSSE spectra \citep{mcconnell98}.  This has been 
largely resolved by a further selection based on the hard X-ray flux.
Figure 1 shows the hard X-ray flux, as measured by the BATSE occultation technique, during 
each of the observations in Table 1.  These data, taken from the BATSE web site 
at Marshall Space Flight Center \footnote{http://www.batse.msfc.nasa.gov/batse/}, 
include statistical errors only.  Also indicated in Figure 1 are the 
$\gamma_0$, $\gamma_1$, and $\gamma_2$ flux levels as defined by \citet{ling87,ling97}.
In general, the hard X-ray flux varied between the $\gamma_1$ and $\gamma_2$ 
levels.
However, the hard X-ray flux was 
considerably below-average (at the $\gamma_0$ level) during Viewing Period (VP) 318.1 and, 
to a lesser extent, during VP 331.5.  Due to the way that the 
data were collected, these low hard X-ray 
flux observations were weighted differently by the various CGRO instruments.
These different weightings were directly responsible for the 
discrepancies noted during our initial analysis.  For the final 
analysis reported here, we excluded the data from VP 318.1 and VP 331.5.
The remaining data corresponded to a relatively constant hard X-ray flux 
of 0.1 photons cm$^{-2}$ s$^{-1}$ in the 45--100 keV energy band.  It is 
assumed, based on the consistent level of hard X-ray flux, that the 
corresponding soft X-ray flux was also consistent during these observations
and that these data all correspond to the canonical low X-ray state of 
Cygnus X-1.  During the first few months of the CGRO mission (from 
the launch in April of 1991 until October of 1991) all-sky monitoring data from 
Ginga (1--20 keV) is available that confirms our assumption that Cygnus X-1
was in its low state \citep{kitamoto2000}.  For the period from October of 
1991 until December 
of 1995 (when RXTE was launched), the data archives at the High Energy 
Astrophysics Science Archive Research Center (HEASARC)
\footnote{http://heasarc.gsfc.nasa.gov/}
show only sporadic pointed observations
by ASCA or ROSAT, none of which correspond precisely to 
any of the COMPTEL observation times.

\section{COMPTEL Data Analysis}

The analysis of COMPTEL data can be logically divided into a spatial
(or imaging) analysis and a spectral analysis.  In practice, however,
the spatial and spectral analysis of the data is inextricably linked
via the PSF and its dependence on
the incident photon spectrum.  The goal of the spatial analysis is to
define, for a given range of photon energy loss values, the
corresponding photon intensity distribution on the sky.  The final
source spectrum is then derived, in a self-consistent manner, from the
results of a spatial analysis in several distinct energy loss
intervals.

\subsection{Event Selections and the COMPTEL Dataspace}

The spatial analysis is performed by first selecting data within some
range of measured (total) energy loss values.  Within the chosen
energy range, events are carefully selected in order to reduce the
background contributons and to maximize the signal-to-noise. These
event selections include the following: a) restrictions on D1 pulse
shape (PSD) to select only those events consistent with incident
photons; b) a TOF selection to reject all but ``forward'' scattered
photon events; c) a scatter angle ($\bar{\phi}$) selection to restrict
the analysis to a range of values which is dominated by source events
rather than by background events; and d) a selection to reject any
event whose event circle passes within $10\arcdeg$ of the Earth's disk
(thus minimizing the background of earth albedo $\gamma$-rays).

Once the event data have been selected, the subsequent analysis is carried
out within a three-dimensional dataspace that is defined by the
fundamental quantities that
represent each COMPTEL event.  The first two of these parameters, the
angles that are arbitarily referred to as $\chi$ and $\psi$, define
the direction of the photon that is scattered from D1 to D2.  The
third parameter defining this dataspace is the Compton scatter angle
($\bar{\phi}$), as estimated from equation (1) using the measured
energy losses in both D1 and D2.

For a point source, the distribution of events in the 3-dimensional
($\chi$, $\psi$, $\bar{\phi}$) dataspace is generally contained
within the interior of a cone whose apex corresponds to the direction
of the source.  This distribution corresponds to the
PSF of COMPTEL.  The details of the PSF depend
not only on the energy range of the analysis, but also on the shape of
the assumed input spectrum, especially as it extends to higher
energies.  In the present analysis, we have derived
PSFs from Monte Carlo simulations \citep{stacy96}.  The current
uncertainties in the PSF are estimated to contribute a systematic
error of not more than $\sim$15--20\% to the flux uncertainties.  This 
error is dominated by uncertainties in the physical modeling of 
COMPTEL and not by Monte Carlo statistics.

\subsection{COMPTEL Spatial Analysis (Imaging)}

The derivation of an image from binned COMPTEL event data (binned into
the three-dimensional dataspace) starts by folding an assumed source
distribution through the instrumental response (PSF).  The PSF
incorporates all of the various physical processes alluded to in $\S$ 2
as well as the various event selections which have been imposed on the
data.  The known exposure and geometric factors (which are specific to
a given observation and include, for example, the on/off status of the
various modules) are then included.  Finally, a `background' model,
which may consist of several components, is added in.  This
process results in an estimate of the event distribution in the
three-dimensional COMPTEL dataspace, which can be compared directly
with the real data.  Subsequent iterations of this process lead to a
more precise estimate of the source distribution on the sky.

In practice, the imaging analysis of the COMPTEL data is performed in
one of two ways.  One technique employs the maximum entropy method to
derive a source distribution on the sky 
\citep{strong92,schonfelder93}.  As presently implemented for COMPTEL data
analysis, this algorithm does not provide quantitative error
estimates.  A more quantitative analysis can be made using a maximum
likelihood technique 
\citep{deboer92,schonfelder93}.
This approach compares the relative probability of a model which
contains only background to the probability of a model which contains
both the background plus a single point source at the given location
(the likelihood ratio).  This method provides quantitative information
regarding both the source location and the flux together with their
associated errors.  

The background model used to generate an image is a crucial component
of the analysis.  These background models consists of components which
describe both the (internal) instrumental background and the
(external) sky background. The term 'background', in this case, refers 
to all photon sources other than the source of interest (Cygnus X-1).
To date, the most successful approach for
estimating the {\em instrumental} background involves a smoothing technique
which suppresses point-source signals while preserving the general
background structure \citep{bloemen94}.  The {\em sky} background
component typically includes both known (or suspected) point sources
and diffuse sources (such as the galactic diffuse emission).  The
maximum likelihood analysis provides a best-fit normalization factor
(and associated uncertainty) for each background component. 

The location of Cygnus X-1 in galactic coordinates ($l = 71.3\arcdeg, 
b = 3.1\arcdeg$) places it in a rather complex region.  
The background modeling therefore included a number of different models 
for the spatial distribution of the celestial photon emission.  These 
models corresponded either to established models for celestial 
$\gamma$-ray emission (e.g., models based on the distribution of 
atomic and molecular gas within the galaxy) or on ad-hoc photon 
distributions as derived directly from COMPTEL images.  Images were 
generated using a variety of such models in various combinations.
In all cases, a point source model at the location of Cygnus X-1 was 
included.  The resulting distribution of flux values for 
the Cygnus X-1 point source model provides a direct measure of the 
errors in the point source analysis, incorporating both the statistical
errors from counting statistics and the systematic errors introduced 
by the background modeling.  

For the purposes of background modeling, there are two known sources of
$\gamma$-ray emission that should be noted.  The first of these is the diffuse
$\gamma$-ray emission from the galaxy. We have made use of a model that is 
consistent with global studies of COMPTEL data 
\citep[e.g.,][]{strong96,bloemen99,bloemen2000}.  This model
includes estimates of the $\gamma$-ray emission which results from the
interaction of cosmic rays with both $HI$ and $H_2$, as well as the contribution 
from inverse Compton emission off cosmic ray electrons \citep{strong95,strong96,strong2000}.  
An $E^{-2}$ power law spectrum is assumed for the diffuse
$\gamma$-ray spectrum.
A second known source of $\gamma$-ray emission is the 39.5 msec 
pulsar PSR 1951+32, which lies only $2.8\arcdeg$ from Cygnus X-1 
(at galactic coordinates $l = 68.8\arcdeg, b = 2.8\arcdeg$). 
Although not readily apparent in spatial studies, a
 timing analysis of COMPTEL data integrated over the full 750 keV to 30 MeV 
energy band independently 
provides evidence for PSR 1951+32 \citep{kuiper98}.  
Its close proximity to Cygnus X-1 makes this source an important 
component of the background models.

A sample of COMPTEL imaging data is shown in Figure 2, where we 
present a maximum likelihood map derived from data integrated over 
the energy loss range of 0.75 -- 2.0 MeV.   
The contours represent constant values of the quantity $-2 \ln{\lambda}$, 
where
$\lambda$ is the likelihood ratio.
In a search for single point sources, $-2 \ln{\lambda}$ has a chi-square
distribution with 3 degrees of freedom.  (For instance, a $3\sigma$ detection
corresponds to $-2 \ln{\lambda}$ = 13.9.)  Cygnus X-1 is clearly 
visible.  The likelihood reaches a value of $-2 \ln{\lambda} =
93.2$ at the position of Cygnus X-1,
which corresponds to a detection significance of $9.7\sigma$.  
These same data were used to derive the $1\sigma$, $2\sigma$ and $3\sigma$ location 
confidence contours shown in Figure 3, which demonstrate the 
ability of COMPTEL to locate the source of emission.  In defining constraints
on the source location, $-2 \ln{\lambda}$ has a chi-square
distribution with 2 degrees of freedom.  So the $1\sigma$, $2\sigma$ and $3\sigma$ location 
confidence contours correspond to a change in likelihood of 2.3, 6.2, and 11.8, 
respectively.
The contours reflect only the statistical uncertainties; 
systematic effects are not included.

\subsection{COMPTEL Spectral Analysis}

A photon spectrum of Cyg X-1 was assembled using flux values derived 
from the spatial (imaging) analysis of COMPTEL data in five distinct energy bands. 
Since the flux measurements are, at some level, dependent on the instrumental PSF 
(and the spectral form assumed for that PSF), the resulting spectrum 
is also dependent on the PSF (and the spectral form assumed for that 
PSF). The analysis therefore included a careful check on the consistency
of the resulting photon spectrum with the spectral form assumed in 
generating each PSF.

Initial fluxes were derived using PSFs based on an $E^{-2}$
power-law spectrum. The resulting photon spectrum was fit with a power-law
spectrum of the form,

\begin{equation}
{dN \over dE} = A E^{-\alpha}
\end{equation}

\noindent
The fit 
gave a photon index of $\alpha = -3.2 (\pm 0.4)$.  This suggested the need to 
use PSFs based on a steeper power-law source spectrum.  Subsequent 
results were 
derived using PSFs based on an $E^{-3}$ power-law spectrum.  A power-law 
fit to this second spectrum gave an index of $\alpha = -3.3 (\pm 0.4)$. 
From this result, we conclude that, at least within this range of PSFs, the resulting 
flux values are relatively insensitive to changes in the PSFs.  In other 
words, the derived photon spectrum is not very compliant with respect to 
the assumed input spectrum.  The robust nature of the extracted photon spectrum 
means that spectral fits in photon dataspace (rather than energy-loss 
dataspace) can be performed with a high degree of reliability.

The final flux data points  from  the analysis of the COMPTEL data are given in Table 2.
These results were derived using PSFs based on an $E^{-3}$ power-law 
spectrum.
The quoted uncertainties include both statistical and systematic errors.
The statistical errors are those due to the counting statistics of the measurement.
The systematic errors are based on results from using various background 
models (as described in $\S$ 4.2) and, in some cases, are
comparable to the statistical errors, especially at lower energies.
(We have not included here the systematic error associated with the PSF 
calculations, as discussed in $\S$ 4.1, because these are considered to be 
negligible relative to other sources of error.)

A plot of the COMPTEL spectra, along with the best-fit power-law fit, is 
shown in Figure 4.  The best-fit power-law gives a 
spectral index 
($\alpha$) of $-3.3 \pm0.4$ and an amplitude of $5.1(\pm0.8) \times 10^{-4}$ cm$^{-2}$ 
s$^{-1}$ MeV$^{-1}$ at 1 MeV.
These data provide convincing evidence for emission up to 
2 MeV, with a marginal detection in the 2--5 MeV energy band.  There 
is no evidence for emission at energies above 5 MeV.  These conclusions
are supported by the images generated from these same data.
The resulting flux values are consistent with previously
published results \citep[e.g.,][]{mcconnell94}.

\section{Broad-Band CGRO Spectral Analysis}

Several different models can be used to fit the COMPTEL data, but the lack of low 
energy data points results in very poor constraints on the model parameters.
We therefore must incorporate additional spectral data (at lower energies) in order to
more precisely define the broad-band spectrum and hence gain some
important insight into the physics of the emission region.
The additional data used in this case are the  contemporaneous
BATSE and OSSE data.  As previously discussed, these data have been 
selected based both on the requirement of contemporaneous observations and 
on the requirement of a consistent level of hard X-ray flux.

For these broad-band spectra, power-law models clearly do not 
provide an adequate description of the data.    We chose four alternative 
spectral forms to help describe the data. The first form is an exponentiated 
power-law,

\begin{equation}
{dN \over dE} = A E^{-\alpha} e^{-E/{E_c}}
\end{equation}

\noindent
This particular function, defined by a power-law index ($\alpha$) and 
a characteristic energy cutoff ($E_c$), is not based on any underlying physical model, but 
its functional form approximates the spectrum at these energies \citep[e.g.,][]{phlips96}.   
The data was also fit using both single-component and two-component 
Comptonization models.
In this case, we used the analytical Comptonization model of \citet{titarchuk94}, which is 
characterized by a normalization
factor, an electron temperature ($kT_e$) and an optical depth ($\tau$).
A fourth functional form is the hybrid thermal/non-thermal model of 
\citet{poutanen96} \citep[see also][]{poutanen98a,poutanen98b}, 
which describes the photon spectrum resulting from a 
thermal (Maxwellian) electron distribution plus a non-thermal (power-law) 
electron distribution at higher energies.  The thermal component is described 
by a characteristic temperature ($kT_e$). The non-thermal component is
characterized as a power law with spectral slope $p_e$ extending from 
an electron Lorentz factor $\gamma_{min}$ (where the Maxwellian transforms to the 
power-law tail) up to a Lorentz factor of 
$\gamma_{max} = 1000$.  The electron population is assumed to reside in a 
accretion disk corona with an optical depth of $\tau$.  This particular model is useful in that 
it permits a quantitative description of the underlying electron distribution.

We have already demonstrated that the reproduced COMPTEL
spectrum is relatively insensitive to the shape of the assumed spectrum,
thus giving us some level of confidence in fitting these data in 
photon-space.  Here we also make the same assumption with regards to both the
BATSE and OSSE spectra.  Given the steepness of the measured spectra, 
this assumption is likely to be a safe one, at least to first-order.
This approach greatly simplifies the analysis using multiple observations
from multiple experiments.

The initial fits to the combined datasets were performed over the 50 keV
to 5 MeV energy range.  A significant amount of scatter in the OSSE and BATSE data 
at energies below 200 keV led to rather poor chi-square values for the resulting 
fits.  In order to obtain improved chi-square values, the final model fits were derived 
from the data between 200 keV and 5 MeV.  In addition to the reduced scatter of 
the BATSE and OSSE data, the general agreement of these two datasets was also 
much improved at energies above 200 keV.  Below 200 keV, there exists discrepancies of up to 25\%
between the BATSE and OSSE data, whereas at energies above 200 keV the 
spectra were found to agree quite closely, with differences typically less 
than 5\%.  
This range is also above the range where many models predict the 
presence of a backscatter component from photons scattering off a 
cooler optically thick accretion disk. 
Spectral fits were derived by allowing both the BATSE and OSSE 
normalizations to be free parameters.   We found that 
constraining the COMPTEL normalization to that of either BATSE or OSSE gave 
consistent results for the physical parameters of interest.  The final fit parameters
were derived by constraining the COMPTEL normalization to that of OSSE.

A summary of the broad-band fit results is given in Table 3.  The broad-band photon spectrum, 
along with the various spectral fits, is shown in Figure 5.  The same data is also 
shown in Figure 6, but now plotted in terms of $E^2$ times the photon flux; 
this plot shows that the power peaks around 100 keV and also accentuates the 
differences between the spectra and the various model fits.
In addition, we also plot (in both Figures 4 and 5) 
an estimated upper limit from EGRET data collected 
during Cycles 1--4 of the CGRO mission, an exposure which is similar to that 
for the other data included in this analysis.  This upper limit is based on data from 
\citet{hartman99} and 
assumes an $E^{-3}$ source spectrum.  
These data do not provide any evidence for a cutoff to the high energy 
power-law.

\section{Discussion}

We have assembled a broad-band $\gamma$-ray spectrum of Cygnus X-1 
using data collected from COMPTEL, 
BATSE and OSSE during the first three years of the CGRO mission.  The data 
were collected contemporaneously and selected so as to have a common 
level of hard X-ray flux.  The hard X-ray flux during these observations
varied between the $\gamma_1$ and $\gamma_2$ levels of \citet{ling87,ling97}.
Although there is poor coverage at soft X-ray energies during this time frame, data from 
both BATSE \citep{ling97} and OSSE {\citep{phlips96} indicate that, for the 
included observations (Table 1), the hard X-ray / $\gamma$-ray spectrum 
remained fairly stable.  In particular, the spectral form is that of the 
``breaking $\gamma$-ray state'', as defined by \citet{grove98}, which is 
generally associated with the low X-ray state.
(\citet{mcconnell2000} compare the spectrum presented here with that 
obtained during a high X-ray state observation in 1996, showing evidence
for spectral variability at MeV energies.)

The COMPTEL data provide evidence of significant emission out to at least 2
MeV. There is additional, but perhaps less compelling, evidence
for emission in the 2--5 MeV energy band, with no  
evidence for emssion above 5 MeV.  An analysis of the COMPTEL data alone does not
adequately constrain the nature of the broad-band  $\gamma$-ray spectrum.  
A more complete interpretation of the COMPTEL data requires the consideration  
of the spectrum at lower energies.  Here we have used contemporaneous data from 
both BATSE and OSSE to define the spectrum at lower energies.

As noted previously \citep{mcconnell94}, these data do not provide any evidence
for an ``MeV bump''.  The spectrum above 750 keV can be described as a power-law 
with a photon spectral index of $\alpha$ = -3.2.  There is no evidence 
for a break in this power-law.  An estimated upper limit from EGRET data 
does not constrain the extrapolated power-law.  An important goal of future 
measurements will therefore be 
the determination of the break energy of this power-law spectrum.

Broad-band spectral fits, at energies above 200 keV, show that both the 
single-component Comptonization model and the exponentiated power-law model do not 
provide a good fit to the data at energies above 1 MeV, although they do 
provide very good fits to the BATSE-OSSE data alone.  The COMPTEL data dictate 
the need for an additional spectral component.  The high energy fit is 
significantly improved using a two-component Comptonization model.  
Unfortunately, the analytical model (T94) begins to fail in this case, because the 
derived electron temperature of the high-temperature component exceeds the 
range that is generally allowed for this model \citep{skibo95a}.   Nonetheless,
the improvements that result from the two-component Comptonization model 
suggest that perhaps a stratified Comptonization region (providing a 
range of both electron temperatures and optical depths) would be more 
appropriate for modeling the spectrum.  \citet{ling97} reached the same
conclusions based on Monte Carlo modeling of BATSE spectra combined with (non-contemporaneous)
COMPTEL data \citep{mcconnell94}.  Physically, this is a much more appealing concept than 
that of a multi-component Comptonization model in which there is no 
interaction between isothermal Comptonization regions.  Thermal gradients are 
incorporated into several models, all of which lead to the generation of 
a high energy tail \citep[e.g.,][]{skibo95b,chakrabarti95,misra96,ling97}.

Hybrid thermal/nonthermal models may also be a viable possibility for explaining
the observed $\gamma$-ray spectrum.  The existence of such distributions is 
clearly established in the case of solar flares \citep[e.g.,][]{coppi99} and it 
is therefore natural to expect that similar distributions exist elsewhere in the 
universe.  Several such models have been discussed 
in the literature \citep[e.g.,][]{crider97,gierlinski97,poutanen96,poutanen98a,poutanen98b,coppi99}. 
Fits to our broad-band spectrum using the model of \citet{poutanen96} provide a 
quantitative estimate of the electron distribution. In this case, the spectral data
can be explained by a thermal Maxwellian distribution with an electron temperature
of $kT_e = 86$ keV, coupled to a power-law electron spectrum with a power-law index of 
$p_e = 4.5$.   The transition between the Maxwellian and the power-law occurs at an 
electron kinetic energy of $\sim570$ keV ($\gamma_{min} = 2.12$).  The electron population is 
embedded in an accretion disk corona with an optical depth of $\tau = 1.63$.  The 
derived spectral index for the power-law tail is somewhat harder (4.5 versus 3.2) but still 
consistent with that derived by \citet{crider97}.

The role of $\pi^o$-production can also not be ruled out as a contribution to the 
measured spectrum.  As initially
proposed by \citet{jourdain94}, this mechanism was used to explain the inadequacy of the
ST80 model at energies below 1 MeV.  It has not been applied to Cygnus
X-1 at energies above 1 MeV, which would be required to test this
model with the present results.  Models for advection-dominated accretion flows 
(ADAF) predict ion temperatures 
of $\sim10^{12}$ K in the inner part of the accretion flow \citep[e.g.,][]{chakrabarti95}, 
suggesting the possibility of pion production (although it is not clear 
whether an  ADAF is present during the low X-ray state of Cygnus X-1).

 The spectrum presented here clearly indicates the need for a non-Maxwellian 
electron energy distribution.  In particular, it strongly suggests the presence 
of a high energy tail to that distribution.  Whether this results from thermal 
gradients or from a more isothermal system coupled with a non-thermal component 
remains an open question.  The shape of the electron distribution  and its 
high energy tail can only be determined by measurements that extend into the MeV  
energy region.

There have been several studies of the broad-band hard X-ray emission from Cygnus 
X-1 based, in part, on OSSE data.  These studies have concentrated on joint 
observations with lower energy experiments, such as GINGA, ASCA or RXTE.  Since 
these observations are of relatively short duration, these spectra typically do 
not exhibit a 
well-defined hard tail due to limited statistics near 1 MeV.  Nonetheless,
the presence of a hard tail in the continuum emission near 1 MeV now appears 
to be well-established.  An excess above the standard Comptonization models
has been discussed several times in the literature for both Cygnus X-1
\citep[e.g.,][]{mcconnell94,phlips96,ling97} and for GRO J0422+32 \citep{vandijk95}.
The data presented here provide the best quantitative measurement 
of the spectrum of Cygnus X-1 at these energies and thus provide the best 
opportunity to study the nature of this hard tail emission. 
The IBIS and SPI instruments on INTEGRAL will 
provide only a marginal improvement in the continuum sensitivity over that 
of CGRO \citep{lichti96,ubertini96}.  In addition, with their much smaller FoV, the INTEGRAL instruments, 
unlike COMPTEL, may be severely limited in terms of their total exposure to 
Cygnus X-1.  COMPTEL may therefore provide the best available data on the MeV 
spectrum of Cygnus X-1 for many years to come.
The continued analysis of data from CGRO (only a fraction of the total 
COMPTEL data are used here) may help to further clarify the nature of this high energy emission 
and perhaps enable us to more fully elucidate the physics of the accretion 
process around stellar-mass black holes.

\acknowledgments

The authors would like to acknowledge Juri Poutanen and Andrzej Zdziarski
for comments on the original manuscript.  Juri Poutanen kindly provided the 
XSPEC version of his hybrid thermal/non-thermal model \citep{poutanen96} that was used 
in this anlaysis.
The COMPTEL project is supported by NASA under contract
NAS5-26645, by the German government through DLR 
grant 50 Q 9096 8 and by the Netherlands Organization for 
Scientific Research NWO.  This work has also been supported at 
UNH by the CGRO Guest Investigator Program under NASA grant 
NAG5-7745.

\begin{figure}
\plotone{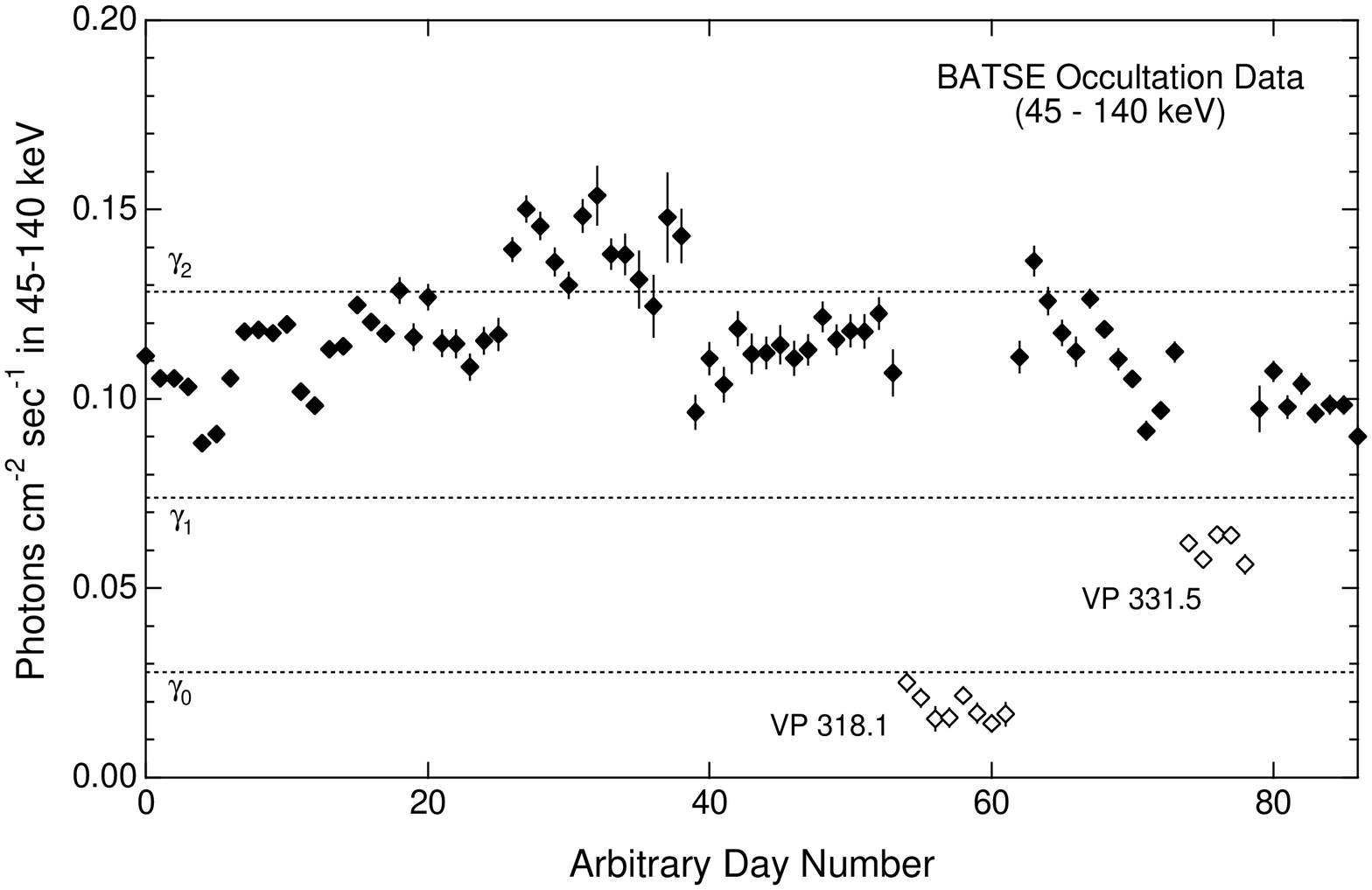}
\caption{Hard X-ray flux measurements from BATSE occultation data 
for those days in which COMPTEL observed Cygnus X-1.
These data cover the CGRO observations listed in Table 1.  The open 
diamonds are from VP 318.1 and VP 331.5, which were excluded from 
the final analysis.  The hard X-ray flux levels defined by \citet{ling87,ling97}
are indicated.}
\end{figure}

\begin{figure}
\plotone{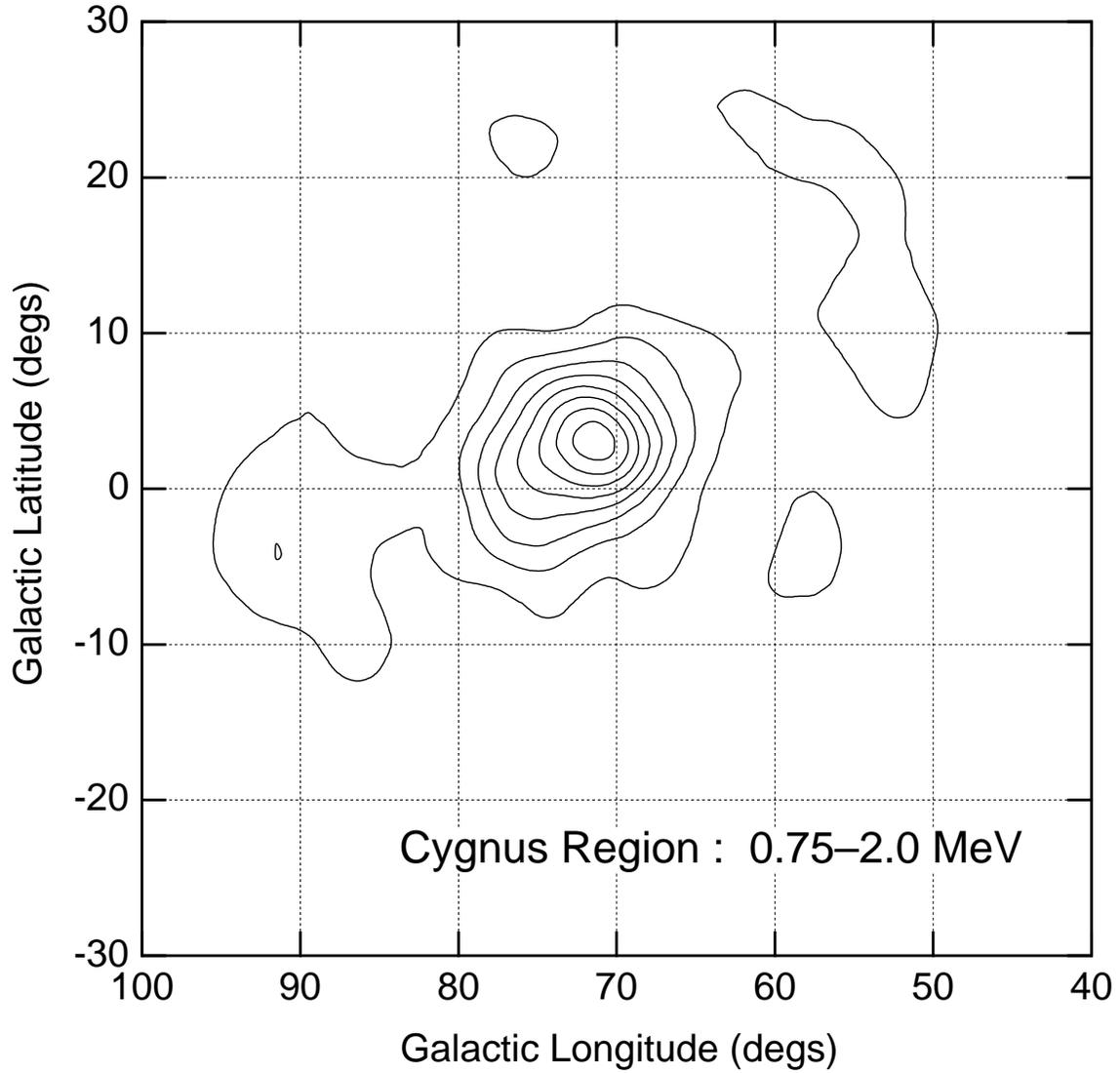}
\caption{COMPTEL maximum likelihood map of the Cygnus region as derived from 
data accumulated over the energy range from 0.75 to 2.0 MeV.
The contour levels start at a likelihood value of 2 with
an increment of 12.}
\end{figure}

\begin{figure}
\plotone{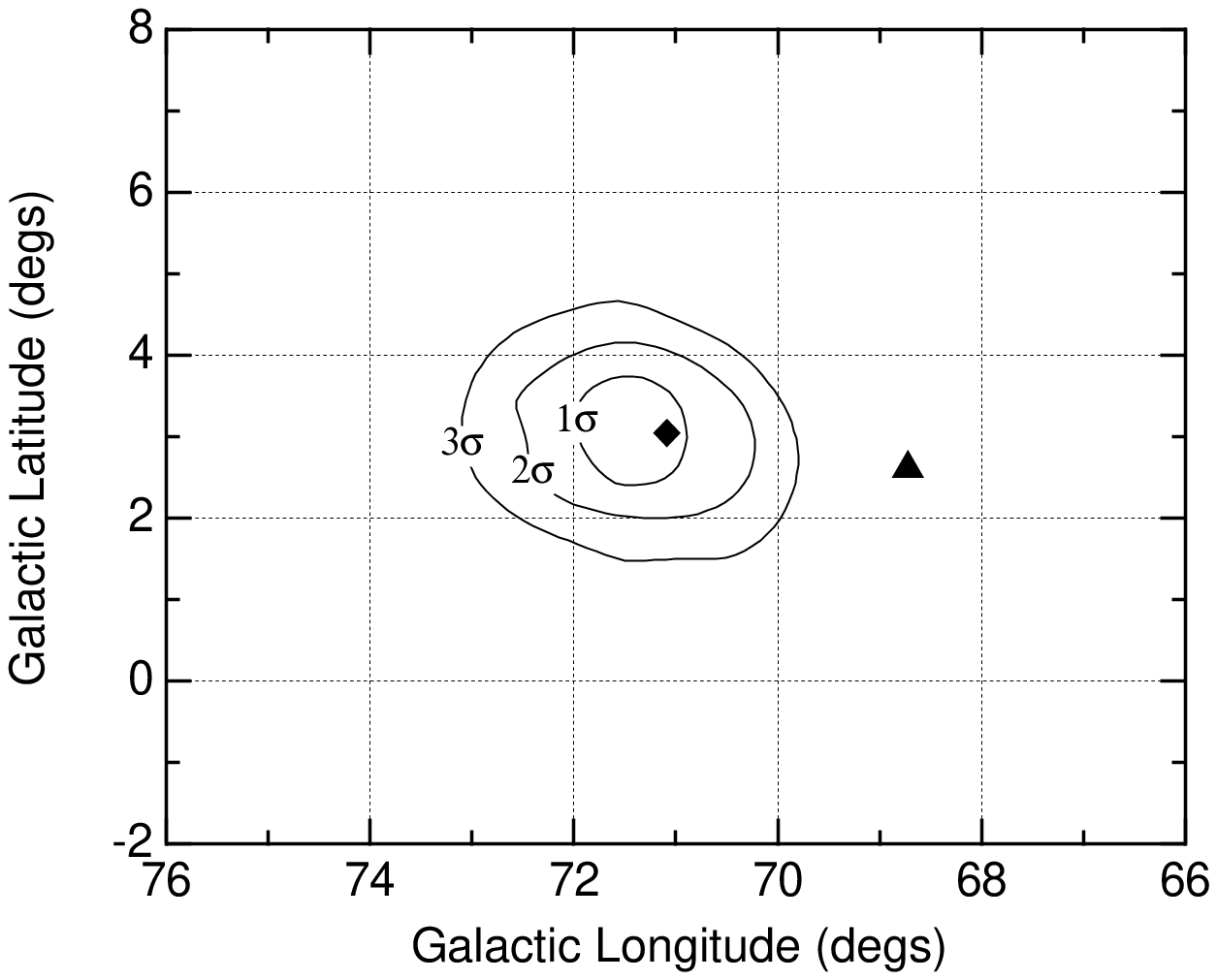}
\caption{COMPTEL location confidence contours for Cygnus X-1 based on the 
maximum likelihood map in Figure 2.  The diamond ($\blacklozenge$) indicates the location 
of Cygnus X-1 ($l = 71.3\arcdeg, b = 3.1\arcdeg$).
The triangle ($\blacktriangle$) indicates the location of PSR1951+32 
($l = 68.8\arcdeg, b = 2.8\arcdeg$), a source
which has been detected in a COMPTEL timing analysis \citep{kuiper98}}.
\end{figure}

\begin{figure}
\plotone{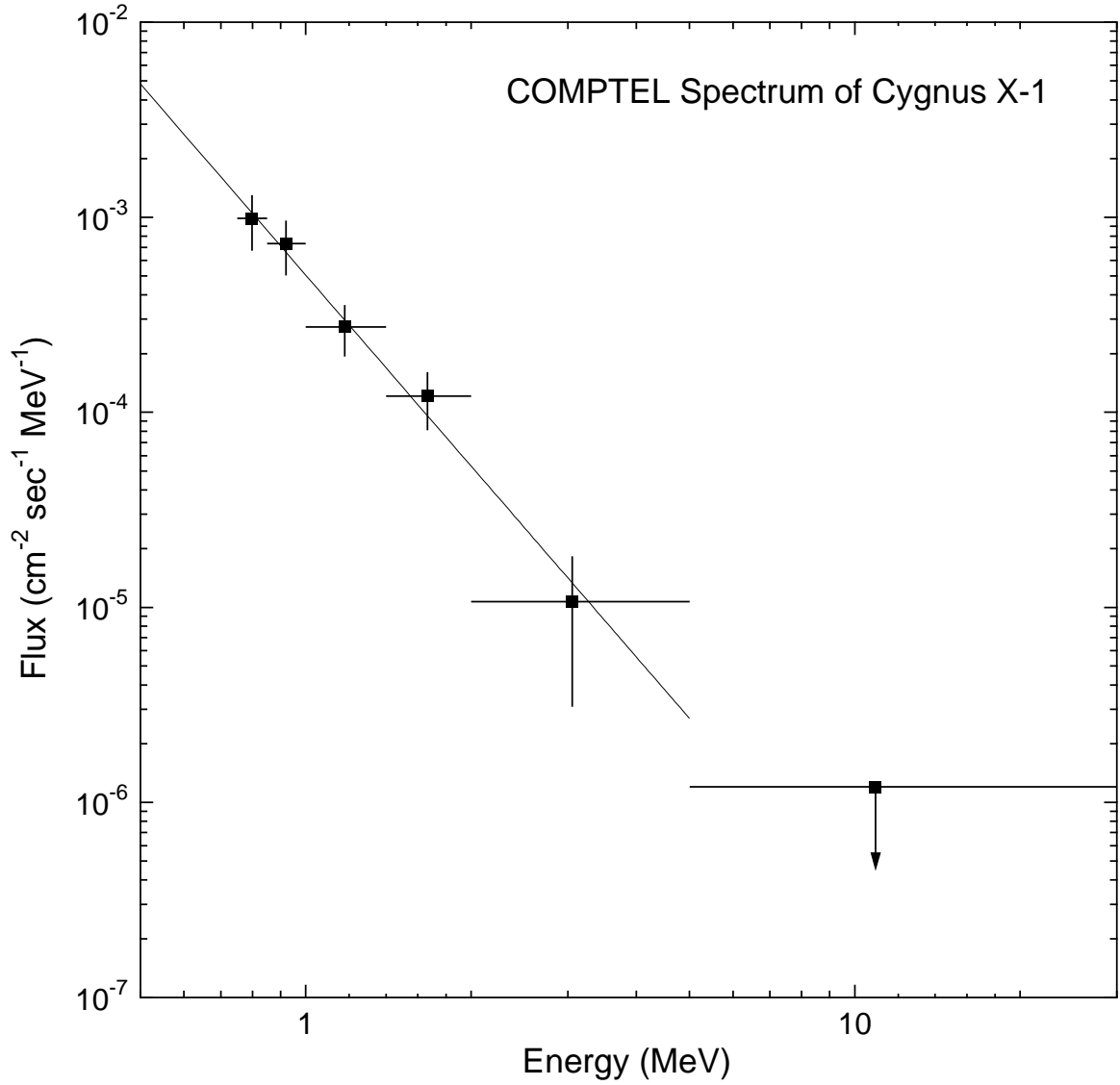}
\caption{The photon spectrum of Cygnus X-1 as derived from COMPTEL data, along with 
the best-fit power-law (with a spectral index of $\alpha = -3.3$).  This spectrum 
corresponds to the data in Table 2.}
\end{figure}

\begin{figure}
\plotone{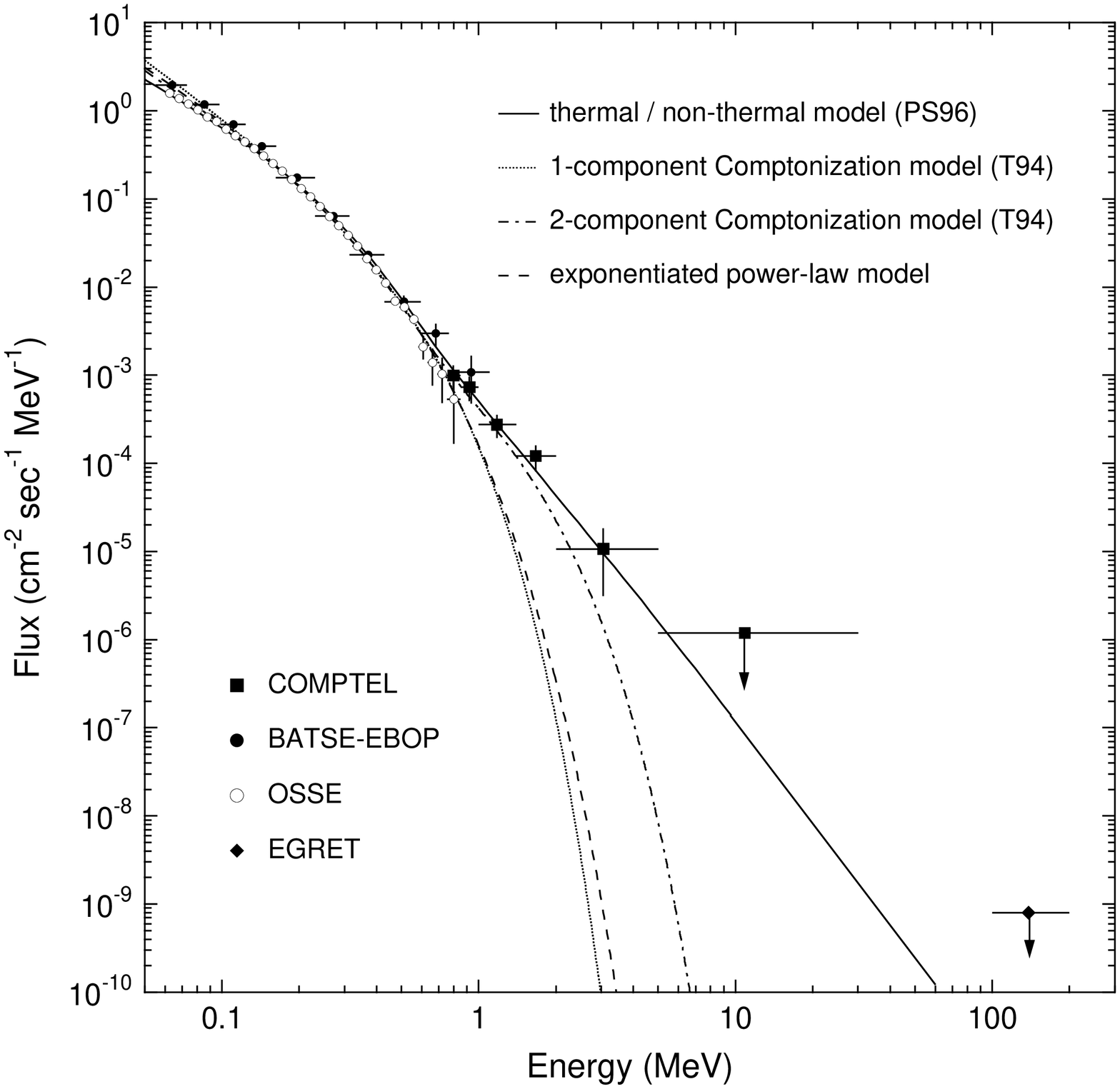}
\caption{Contemporaneous broad-band spectrum of Cygnus X-1 including data 
from COMPTEL, OSSE and BATSE.  Also shown are the various model fits, whose 
parameters are listed in Table 3.  The data are plotted in units of photon flux.
 For the sake of clarity, upper limits from 
OSSE and BATSE are not shown, but these are consistent with the total 
dataset. The EGRET upper limit is based on data from \citet{hartman99}.}
\end{figure}

\begin{figure}
\plotone{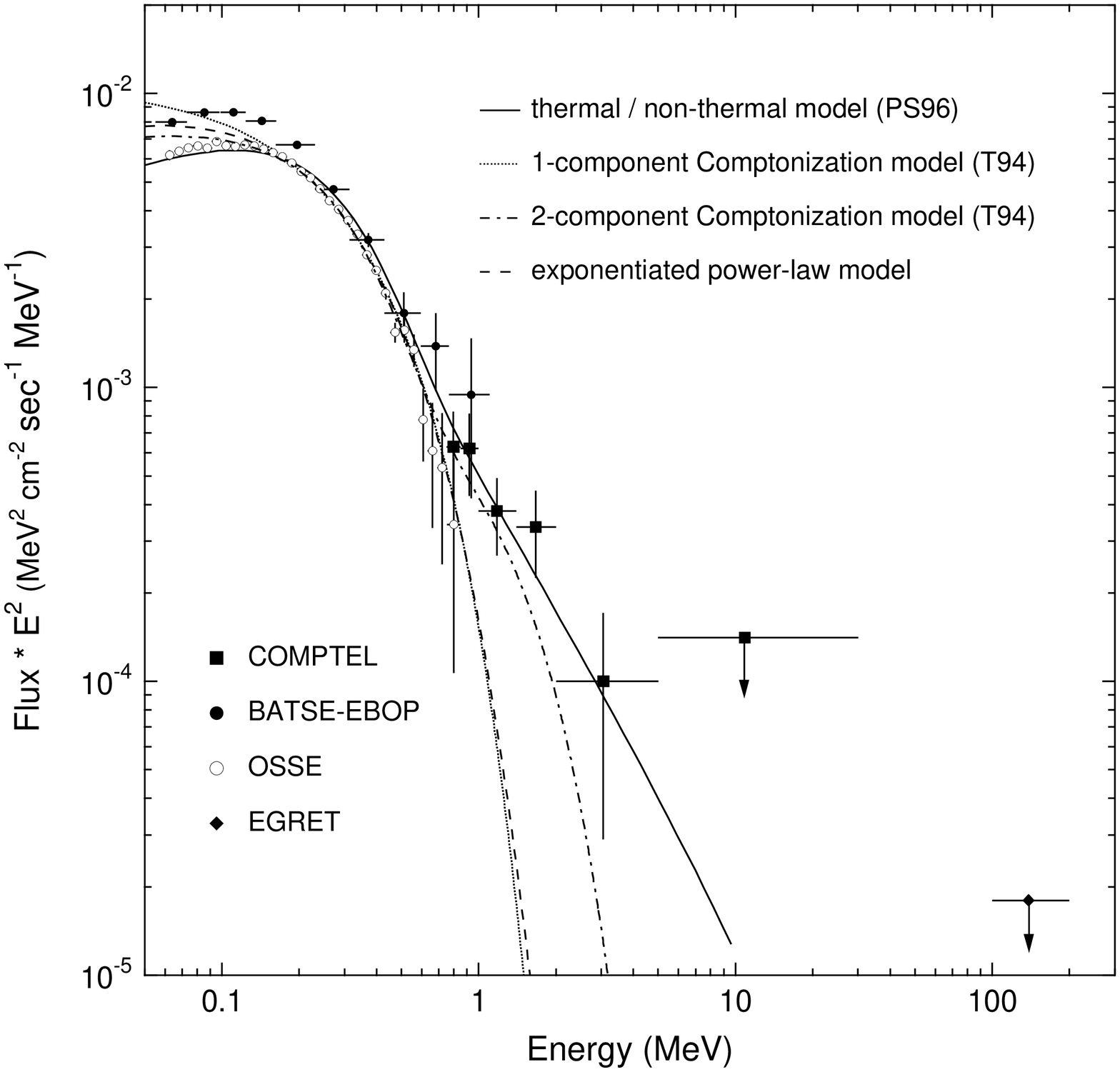}
\caption{Contemporaneous broad-band spectrum of Cygnus X-1 including data 
from COMPTEL, OSSE and BATSE.  Also shown are the various model fits, whose 
parameters are listed in Table 3.  The data are plotted in units of $E^2$ times 
the photon flux.
 For the sake of clarity, upper limits from 
OSSE and BATSE are not shown, but these are consistent with the total 
dataset. The EGRET upper limit is based on data from \citet{hartman99}.}
\end{figure}

\begin{deluxetable}{cccccccc} 
\tablewidth{0pt}
\tablehead{\colhead{Viewing} & \colhead{Start} & \colhead{Start} & \colhead{End} & \colhead{End} & \colhead{Viewing}
 & \colhead{Effective} & \colhead{Used in} \\
\colhead{Period} & \colhead{Date} & \colhead{TJD} & \colhead{Date} & \colhead{TJD} & \colhead{Angle}
 & \colhead{Exposure\tablenotemark{a}} & \colhead{Analysis?}}
\tablecaption{Relevant CGRO Viewing Periods}
\startdata
        &              &        &              &         &               &       &      \\
2.0     & 30-May-1991  & 8406   &  8-Jun-1991  &  8415   &  $1.7^\circ$  & 3.65  & yes  \\
7.0     &  8-Aug-1991  & 8476   & 15-Aug-1991  &  8483   & $11.2^\circ$  & 2.72  & yes  \\
203.0   &  1-Dec-1992  & 8957   &  8-Dec-1992  &  8964   &  $7.0^\circ$  & 1.75  & yes  \\
203.3   &  8-Dec-1992  & 8964   & 15-Dec-1992  &  8971   &  $7.0^\circ$  & 1.75  & yes  \\
203.6   & 15-Dec-1992  & 8971   & 22-Dec-1992  &  8978   &  $7.0^\circ$  & 1.69  & yes  \\
212.0   &  9-Mar-1993  & 9055   & 23-Mar-1993  &  9069   & $15.4^\circ$  & 2.71  & yes  \\
318.1   &  1-Feb-1994  & 9384   &  8-Feb-1994  &  9391   &  $4.5^\circ$  & 1.78  &  no  \\
328.0   & 24-May-1994  & 9496   & 31-May-1994  &  9503   &  $7.0^\circ$  & 1.56  & yes  \\
331.0   &  7-Jun-1994  & 9510   & 10-Jun-1994  &  9513   &  $7.0^\circ$  & 0.95  & yes  \\
331.5   & 14-Jun-1994  & 9517   & 18-Jun-1994  &  9521   &  $7.0^\circ$  & 1.34  &  no  \\
333.0   &  5-Jul-1994  & 9538   & 12-Jul-1994  &  9545   &  $7.0^\circ$  & 1.86  & yes  \\
        &              &        &              &         &               &       &      \\
\enddata
\tablenotetext{a}{Effective on-axis exposure, measured in days.}
\end{deluxetable}

\begin{deluxetable}{ccccc}
\tablewidth{0pt}
\tablehead{
 \colhead{Energy}             &  \colhead{Counts in}   & \colhead{Source}      & \colhead{Flux}                   \\
\colhead{(MeV)}  &  \colhead{Dataspace}   & \colhead{Counts}      & \colhead{(cm$^{-2}$ s$^{-1}$ MeV$^{-1}$)}  
}
\tablecaption{COMPTEL Flux Measurements for Cygnus X-1.  }
\startdata                 
0.75 -- 0.85  &    63,507 & $ 943\pm298$ & $9.8(\pm3.1) \times 10^{-4}$  \\
0.85 -- 1.0   &   145,197 & $1807\pm572$ & $7.2(\pm2.3) \times 10^{-4}$  \\
1.0 -- 1.4    &   372,823 & $2619\pm775$ & $2.7(\pm0.8) \times 10^{-4}$   \\
1.4 -- 2.0    &   452,226 & $2281\pm760$ & $1.2(\pm0.4) \times 10^{-4}$  \\
2.0 -- 5.0    &   829,813 & $1186\pm835$ & $1.1(\pm0.8) \times 10^{-5}$  \\
5.0 -- 30.0   &   219,916 & $ 331\pm345$ & $3.9(\pm4.0) \times 10^{-7}$  \\
\enddata
\end{deluxetable}

\begin{deluxetable}{lccc}
\tablecolumns{2}
\tabletypesize{\small}
\tablewidth{0pt}
\tablecaption{Spectral Fit Summary (200 keV -- 5 MeV)}
\tablehead{}
\tablecolumns{2}
\startdata
\cutinhead{Exponential Power-Law Model (with 90\% confidence limits)}
$\alpha$                            & $1.71^{+.26}_{-.27}$  \\
$E_c$ (keV)                         & $200^{+37}_{-32}$        \\
Red Chi-Square ($\chi^{2}_{\nu}$)   & 1.53                  \\
Degrees of Freedom ($\nu$)          & 23                       \\[6pt]
\cutinhead{One-Component Comptonization Model (with 90\% confidence limits)}
$kT$ (keV)                          & $ 127^{+45}_{-25}$               \\
$\tau$                              & $0.32^{+.18}_{-.16}$   \\
Red Chi-Square ($\chi^{2}_{\nu}$)   & 1.67                             \\
Degrees of Freedom ($\nu$)          & 23                            \\[6pt]
\cutinhead{Two-Component Comptonization Model (best-fit parameters only)}
$kT_{1}$ (keV)                      & 340          \\
$\tau_{1}$                          & 0.30          \\
$kT_{2}$ (keV)                      & 72              \\
$\tau_{2}$                          & 0.81         \\
$A_1 / A_2$                         & .012     \\
Red Chi-Square ($\chi^{2}_{\nu}$)   & 0.90        \\
Degrees of Freedom ($\nu$)          & 19             \\
\cutinhead{Hybrid Thermal / Non-Thermal Model (with 90\% confidence limits)}
$kT_{e}$ (keV)                      & $86^{+18}_{-8}$           \\
$p_e$                               & $4.50^{+.82}_{-.96}$         \\
$\gamma_{min}$                      & $2.12^{+.79}_{-.49}$         \\
$\tau$                              & $1.63^{+1.15}_{-.57}$           \\
Red Chi-Square ($\chi^{2}_{\nu}$)   & 0.83         \\
Degrees of Freedom ($\nu$)          & 21                \\
\enddata

\end{deluxetable}


\begin{thebibliography}{}

\bibitem[Bednarek et al.(1990)]{bednarek90} Bednarek, W., et al. 1990, \aap, 236,175

\bibitem[Bloemen et al.(1994)]{bloemen94} Bloemen, H., et al. 1994, \apjs, 92, 419

\bibitem[Bloemen et al.(1999)]{bloemen99} Bloemen, H., et al., 1999, Astrophys. Lett. \& Comm., 39, 205

\bibitem[Bloemen et al.(2000)]{bloemen2000} Bloemen, H., et al. 2000, in Proceedings of the 
Fifth Compton Symposium (AIP Conf. Proc. 510), ed. M. L. McConnell \& J. M. Ryan
(New York: American Institute of Physics), in press

\bibitem[de Boer et al.(1992)]{deboer92} de Boer, H., et al. 1992, Proc. 4th International Workshop on 'Data 
Analysis in Astronomy', Erice, Italy (Plenum Press), p. 241

\bibitem[Chakrabarti \& Titarchuk(1995)]{chakrabarti95} Chakrabarti, S. K., \& Titarchuk, L. G.  1995, \apj, 455, 623

\bibitem[Crider et al.(1997)]{crider97} Crider, A., Liang, E. P., Smith, I. A., Lin, D., \& Kusnose, M. 1997, 
in Proceedings of the Fourth Compton Symposium (AIP Conf. Proc. 410), ed. C. D. Dermer, M. S. Strickman, \& J.D. Kurfess
(New York: American Institute of Physics), p. 868

\bibitem[Coppi(1999)]{coppi99} Coppi, P.S.  1999, in High Energy Processes in Accreting Black Holes (ASP Conf. Ser. 161),
ed. J. Poutanen \& R. Svensson (San Francisco: ASP), p. 375.

\bibitem[Dahlbacka et al.(1974)]{dahlbacka74} Dahlbacka, G. H., Chapline, G. F., \& Weaver, T. A. 1974, Nature, 
250, 36

\bibitem[Dermer \& Liang(1989)]{dermer89} Dermer, C. D., \& Liang, E. P.  1989, \apj, 339, 512

\bibitem[Dermer, Miller, \& Li(1996)]{dermer96} Dermer, C. D., Miller, J. A., \& Li H.  1996, \apj, 456, 106.

\bibitem[van Dijk et al.(1995)]{vandijk95} van Dijk, R., et al.  1995, \aap, 296, L33

\bibitem[Dove, Wilms \& Begelman(1997)]{dove97a} Dove, J. B., Wilms, J., \& Begelman, M. C.  1997, \apj, 487, 747

\bibitem[Dove et al.(1997)]{dove97b} Dove, J. B., et al. 1997, \apj, 487, 759

\bibitem[Ebisawa et al.(1996)]{ebisawa96} Ebisawa, K., Ueda, Y., Inoue, H., Tanaka, Y., \& White, N. E.  1996, \apj, 467, 419

\bibitem[Eilik(1980)]{eilik80} Eilik, J. 1980, \apj, 236, 664

\bibitem[Eilik \& Kafatos(1983)]{eilik83} Eilik, J., \& Kafatos, M.  1983, \apj, 271, 804

\bibitem[Gierlinski et al.(1997)]{gierlinski97} Gierlinski, M., Zdziarski, A. A., Done, C., Johnson, W. N., 
Ebisawa, K., Ueda, Y., Haardt, F., \& Phlips, B. F.  1997, \mnras, 288, 958

\bibitem[Grabelsky et al.(1993)]{grabelsky93} Grabelsky, D. A., et al.  1993, in Compton Gamma-Ray 
Observatory (AIP Conf. Proc. 280), ed. M. Friedlander, N. Gehrels, \& D. J. Macomb (New York; American 
Institute of Physics), p. 345

\bibitem[Grebenev et al.(1993)]{grebenev93} Grebenev, S., et al. 1993, \aaps, 97, 281

\bibitem[Grove et al.(1998)]{grove98} Grove, J. E., Johnson, W. N., Kroeger, R. A., McNaron-Brown, K., 
Skibo, J. G., \& Phlips, B. F.  1998, \apj, 500, 899

\bibitem[Haardt et al.(1993)]{haardt93} Haardt, F., Done, C., Matt, G., \& Fabian, A. C. 1993, \apjl, 
411, L95

\bibitem[Harmon et al.(1992)]{harmon92} Harmon, B. A., et al. 1992, in NASA CP-3137, Compton Observatory Science Workshop,
 ed. C. R. Shrader, N. Gehrels, \& B. Dennis (Washinton, DC: NASA), 69

\bibitem[Harris et al.(1993)]{harris93} Harris, M. J., Share, G. H., Leising, M. D., \& Grove, J. E.  1993, \apj, 416, 601

\bibitem[Hartman et al.(1999)]{hartman99} Hartman, R. C., et al.  1999, \apjs, 123, 79

\bibitem[Hua \& Titarchuk(1995)]{hua95} Hua, X. M., \& Titarchuk, L.  1995, \apj, 449, 188

\bibitem[Jordain \& Roques(1994)]{jourdain94} Jourdain, E., \& Roques, J. P.  1994, \apjl, 426, L11

\bibitem[Kitamoto et al.(2000)]{kitamoto2000} Kitamoto, S., Egoshi, W., Miyamoto, S., Tsunemi, H., Ling, J. C.,  
Wheaton, W. A., \& Paul, B.  2000, \apj, 531, 546

\bibitem[Kuiper et al.(1998)]{kuiper98} Kuiper, L., Hermsen, W., Bennett, K., Carrami\~nana, A., 
McConnell, M., \& Sch\"onfelder, V.  1998, \aap, 337, 421

\bibitem[Li \& Miller(1997)]{li97} Li, H., \& Miller, J. A.  1997, \apjl, 478, L67

\bibitem[Li, Kusunose \& Liang(1996)]{li96} Li, H., Kusunose, M., \& Liang, E. P.  1996, \apjl, 460, L29

\bibitem[Liang \& Nolan(1983)]{liang83} Liang, E. P., \& Nolan, P. L. 1983, \ssr, 38, 353

\bibitem[Liang \& Dermer(1988)]{liang88} Liang, E. P., \& Dermer, C. D. 1988, \apjl, 325, L39  

\bibitem[Lichti et al.(1996)]{lichti96} Lichti, G. L., et al. 1996, Proc. SPIE, 2806, 217

\bibitem[Ling et al.(1983)]{ling83} Ling, J. C., Mahoney, W. A., Wheaton, W. A., Jacobsen, A. S., 
\& Kaluzienski, L.  1983, \apj, 275, 307

\bibitem[Ling et al.(1987)]{ling87} Ling, J. C., Mahoney, W. A., Wheaton, W. A., \& Jacobsen, A. S., 
 1987, \apjl, 321, L117
 
\bibitem[Ling \& Wheaton(1989)]{ling89} Ling, J. C., \& Wheaton, W. A. 1989, \apjl, 343, L57

\bibitem[Ling et al.(1996)]{ling96} Ling, J. C., Wheaton, W. A., Mahoney, W. A., Skelton, R. T., 
Radocinski, R. G., and Wallyn, P.  1996, \aaps, 120, C677

\bibitem[Ling et al.(1997)]{ling97} Ling, J. C., et al.  1997, \apj, 484, 375

\bibitem[Ling et al.(2000)]{ling2000} Ling, J. C.,  et al., 2000, \apjs, 127, 79

\bibitem[McConnell et al.(1989)]{mcconnell89} McConnell, M. L., et al. 1989, \apj, 343, 317

\bibitem[McConnell et al.(1994)]{mcconnell94} McConnell, M. L., et al.  1994, \apj, 424, 933

\bibitem[McConnell et al.(1998)]{mcconnell98} McConnell, M. L. et al.  1998, in Proceedings of the 
Fourth Compton Symposium (AIP Conf. Proc. 410), ed. C. D. Dermer, M. S. Strickman \& J. D. Kurfess
(New York: American Institute of Physics), p. 829

\bibitem[McConnell et al.(2000)]{mcconnell2000} McConnell, M. L. et al.  2000, in Proceedings of the 
Fifth Compton Symposium (AIP Conf. Proc. 510), ed. M. L. McConnell \& J. M. Ryan
(New York: American Institute of Physics), in press

\bibitem[Melia \& Misra(1993)]{melia93} Melia, F., \& Misra, R.  1993, \apj, 411, 797

\bibitem[Misra \& Melia(1996)]{misra96} Misra, R., \& Melia, F. 1996, \apj, 467, 405

\bibitem[Misra et al.(1997)]{misra97} Misra, R., Chitnis, V. R., Melia, F., \& Rao, A. R. 1997, \apj, 487, 388

\bibitem[Misra et al.(1998)]{misra98} Misra, R., Chitnis, V. R., \& Melia, F.  1998, \apj, 495, 407

\bibitem[Moskalenko, Collmar \& Sch\"onfelder(1998)]{moskalenko98} Moskalenko, I. V., Collmar, W., \& Sch\"onfelder, V. 1998, \apj, 428, 436

\bibitem[Nolan et al.(1981)]{nolan81} Nolan, P., et al. 1981, Nature, 293, 275

\bibitem[Nolan \& Matteson(1983)]{nolan83} Nolan, P., \& Matteson, J. L.  1983, \apj, 265, 389

\bibitem[Owens \& McConnell(1992)]{owens92} Owens, A., \& McConnell, M.  1992, Comments Astrophys., 16, 205

\bibitem[Phlips et al.(1996)]{phlips96} Phlips, B., et al. 1996, \apj, 465, 907

\bibitem[Poutanen \& Svensson(1996)]{poutanen96} Poutanen, J. \& Svensson, R.  1996, \apj, 470, 249  (PS96)

\bibitem[Poutanen(1998)]{poutanen98a} Poutanen, J. 1998, in Theory of Black Hole Accretion Disks, eds. M. A.
Abramowicz, G. Bj\"ornsson \& J. E. Pringle (Cambridge Univ. Press: New York), p. 100

\bibitem[Poutanen \& Coppi(1998)]{poutanen98b} Poutanen, J., \& Coppi, P.  1998, Physica Scripta, T77, 57

\bibitem[Poutanen(2000)]{poutanen2000} Poutanen, J.  2000, private communications.

\bibitem[Priedhorsky, Terrell \& Holt(1983)]{priedhorsky83} Priedhorsky, W. C., Terrell, J., \& Holt, S. S.  1983, \apj, 270, 233

\bibitem[Salotti et al.(1992)]{salotti92} Salotti, L., et al. 1992, \aap, 253, 145

\bibitem[Sch\"onfelder \& Lichti(1974)]{schonfelder74} Sch\"onfelder, V., \& Lichti, G.  1974, \apjl, 192, L1

\bibitem[Sch\"onfelder et al.(1993)]{schonfelder93} Sch\"onfelder, V., et al.  1993, \apjs, 86, 629

\bibitem[Shapiro, Lightman and Eardley(1976)]{shapiro76} Shapiro, S. L., Lightman, A. P., \& Eardley, D. M. 1976, \apj, 
204, 187

\bibitem[Skibo et al.(1995)]{skibo95a} Skibo, J. G., Dermer, C. D., Ramaty, R., \& McKinley, J. M.  1995, \apj, 446, 86

\bibitem[Skibo \& Dermer(1995)]{skibo95b} Skibo, J. G., \& Dermer, C. D.  1995, \apjl, 455, L25

\bibitem[Stacy et al.(1996)]{stacy96} Stacy, J. G., et al.  1996, \aaps, C120, 691

\bibitem[Steinle et al.(1982)]{steinle82} Steinle, H., et al. 1982, \aap, 107, 350

\bibitem[Strong et al.(1992)]{strong92} Strong, A. W., et al. 1992, Proc. 4th International Workshop on 'Data 
Analysis in Astronomy', Erice, Italy (Plenum Press), p. 251

\bibitem[Strong \& Youseffi(1995)]{strong95} Strong, A. W., \& Youseffi, G.  1995, Proc. 24th Internat. Cosmic Ray 
Conf. (Durban), 3, 48

\bibitem[Strong et al.(1996)]{strong96} Strong, A. W., et al. 1996, \aaps, 120, C381

\bibitem[Strong, Moskalenko, \& Reimer(2000)]{strong2000} Strong, A. W., Moskalenko, I. V., and 
Reimer, O.  2000, in Proceedings of the 
Fifth Compton Symposium (AIP Conf. Proc. 510), ed. M. L. McConnell \& J. M. Ryan
(New York: American Institute of Physics), in press

\bibitem[Sunyaev \& Titarchuk(1980)]{sunyaev80} Sunyaev, R. A., \& Titarchuk, L. G.  1980, \aap, 86, 121  (ST80)

\bibitem[Sunyaev \& Tr\"umper(1979)]{sunyaev79} Sunyaev, R. A., \& Tr\"umper, J.  1979, Nature, 279, 506

\bibitem[Titarchuk(1994)]{titarchuk94} Titarchuk, L.  1994, \apj, 434, 570  (T94)

\bibitem[Titarchuk \& Lyubarskij(1995)]{titarchuk95a} Titarchuk, L., \& Lyubarskij  1995, \apj, 450, 876

\bibitem[Titarchuk \& Hua(1995)]{titarchuk95b} Titarchuk, L., \& Hua, X.-M.  1995, \apj, 452, 226

\bibitem[Ubertini et al.(1991a)]{ubertini91a} Ubertini, P., et al. 1991a, \apj, 366, 544

\bibitem[Ubertini et al.(1996)]{ubertini96} Ubertini, P., et al.  1996, Proc. SPIE, 2806, 246

\bibitem[Ubertini et al.(1991b)]{ubertini91b} Ubertini, P., Bazzano, A., La Padula, C., Manchada, R. K., 
Polcaro, V. F., Staubert, R., Kendziorra, E., \& Perotti, F.  1991b, \apj, 383, 263

\bibitem[Winkler et al.(1992)]{winkler92} Winkler, C. et al. 1992, \aap, 255, L9

\bibitem[Zdziarski(1984)]{zdziarski84} Zdziarski, A. A.  1984, \apj, 283, 842

\bibitem[Zdziarski(1985)]{zdziarski85} Zdziarski, A. A.  1985, \apj, 289, 514

\bibitem[Zdziarski(1986)]{zdziarski86} Zdziarski, A. A.  1986, \apj, 303, 94

\end{thebibliography}
\end{document}